\documentstyle[prd,aps,twocolumn,psfig]{revtex}

\newcommand{\beq}{\begin{equation}}
\newcommand{\beqa}{\begin{eqnarray}}
\newcommand{\eeq}{\end{equation}}
\newcommand{\eeqa}{\end{eqnarray}}
\newcommand{\p}{\phi}
\newcommand{\k}{\kappa}
\newcommand{\non}{\nonumber}

\begin{document}
\twocolumn[\hsize\textwidth\columnwidth\hsize\csname
@twocolumnfalse\endcsname
\draft
\title{
Extended Open Inflation
}
\author{
Takeshi Chiba and Masahide Yamaguchi
}
\address{
Department of Physics, University of Tokyo,
Tokyo 113-0033, Japan
}
\date{\today}
\maketitle

\begin{abstract}
We present a new type of one-field models for open inflation 
utilizing a nonminimally coupled  scalar field with polynomial
potentials in which the Coleman-de Luccia instanton does exist and 
slow-roll inflation after the bubble nucleation is realized.

\end{abstract}

]


The possibility of the creation of an open universe within the context
of the inflationary scenario\cite{gott,open1,open2,open3} is
intriguing in that it enriches predictions of inflationary cosmology
and that the horizon problem and the flatness problem can be solved
separately within the inflationary paradigm.  The basic idea is that a
symmetric bubble nucleates in de Sitter space\cite{cd,gott} and its
interior undergoes a second stage of slow-roll inflation adjusting
$\Omega_0$ in the rage $0 < \Omega_0 <1$.

As noted by Linde, however, it is very difficult to satisfy both
$|V''|/H^2> 1$ (the condition for the existence of Coleman-de Luccia
instantons) and $ |V''|/H^2\ll 1$ (for slow-roll inflation after the
bubble nucleation)\cite{linde}.  In fact, we did not have any
consistent models until Linde proposed a simple one-field
model\cite{linde}(see also \cite{lst}).  The purpose of the present
paper is to propose some version of open inflation satisfying both
requirements using a nonminimally coupled scalar field with 
polynomial potentials. We shall name our model ``extended'' open
inflation in the sense of \cite{ls}. The possibility of open inflation
with a nonminimally coupled scalar field has been
suggested in \cite{gott2}. 
However, as far as we know,  the concrete
realization has not appeared in the literature\footnote{In fact, in
  \cite{linde}, Linde noted ``... We also tried to use scalar fields
  nonminimally coupled to gravity. All these attempts so far did not
  lead to a successful one-field open universe scenario''.}
(we note that variants of the Hawking-Turok instanton for
  creation of an open universe\cite{ht} with a nonminimally coupled
  scalar field were considered in \cite{bar}).

Linde's potential consists of two parts:
\beqa
V(\phi)&=&{1\over 2}m^2\phi^2+{1\over 2}m^2\phi^2{1\over
  {\beta^2/\alpha^2}+(\phi-v)^2/\alpha^2} \non \\
     &\equiv&
V_1(\phi)+V_2(\phi){1\over f(\phi)^2},
\label{linde}
\eeqa
where the first term, $V_1(\phi)$, controls inflation after quantum
tunneling, while the second term, $V_2(\phi)/f(\phi)^2$, controls the
bubble nucleation. The essential role of $f(\phi)$ is to create a dip
in the potential $V(\phi)$ around the minimum of $f(\phi)$.  Somewhat
{\it ad hoc} feature of this model is the appearance of the second
term in which a polynomial potential $V_2(\phi)$ is divided by another
polynomial function $f(\phi)^{2}$. Thus if the potential of the form
$f(\phi)^2V(\phi)$ appears at the very introduction of the action and
the potential $V(\phi)$ appears as an {\it effective} potential, then
we may avoid the unnaturalness to some extent. These considerations
suggest a nonminimal coupling of the scalar field to gravity in the
following form:
\beqa
S&=&\int d^4x\sqrt{-g}\left[{1\over 2\k^2}f(\phi)R-{1\over
  2}g^{ab}\partial_a\p\partial_b\p \right. \non \\
 && \left. -f(\phi)^2 V_1(\phi)-V_2(\phi)
\right],
\label{action}
\eeqa
where $\k^2=8\pi G_*$ and $G_*$ is the bare gravitational constant.
Indeed after the conformal transformation of the form
\beq
g_{Eab}=f(\phi)g_{ab},
\eeq
the action Eq.(\ref{action}) reduces that of the scalar field
minimally coupled to the Einstein gravity:
\beqa
S&=&\int d^4x\sqrt{-{g_E}}\left[{1\over 2\k^2}R_E-
{1\over 2}F(\phi)^2({\nabla}_E\phi)^2 \right. \non \\
 && \left. -V_E(\phi)\right],
\label{e-action}
\eeqa
where
\beqa
F(\phi)^2&=& {1\over f(\phi)^{2}}\left(f(\phi)+{3f(\phi)_{,\phi}^{~~2}\over
    2\k^2}\right),\\
V_E(\phi)&=&V_1(\phi)+V_2(\phi){1\over f(\phi)^2}.
\eeqa
Here the subscript represents the functional derivative with respect
to $\phi$.  Apart from the non-canonical kinetic term, the potential
$V(\phi)$ in Eq.(\ref{e-action}) has essentially the same feature as
Linde's model given in Eq.(\ref{linde}).

The appearance of the potential terms of different coupling to $\phi$
may seem rather {\it ad hoc}. We speculate on a possible origin of
these potentials. One possibility is a massive field 
with a potential of the form $V_{1}(\phi)$ in the string
frame\cite{dilaton}. Hence we may regard the Lagrangian
Eq.(\ref{action}) without $V_{2}(\phi)$ as a starting point. Another 
potential $V_{2}(\phi)$ may come from a nonperturbative effect
associated with a dynamical symmetry breaking.

An $O(4)$-symmetric instanton with the metric (in the original frame
$g_{ab}$)
\beq
ds^2=d\tau^2+a(\tau)^2(d\psi^2+\sin^2\psi d\Omega_2^2),
\eeq
which describes the creation of an open universe should satisfy the
following equations of motion
\beqa
{a''\over a}&=&-{\k^2\over 3f}\left(\phi'^2+V\right)
-{f_{,\phi} \over 2f}{a'\over a}\phi'-{f_{,\phi\phi}\over 2f}\phi'^2 -
{f_{,\phi}\over 2f}\phi'',\\
\phi''&=&-3{a'\over a}\phi'+ V_{,\phi} +{3 f_{,\phi} \over \k^2}
\left({a''\over a}+\left({a'\over a}\right)^2-{1\over a^2}
\right),
\eeqa
where primes denote derivatives with respect to $\tau$. A regular
instanton should satisfy the boundary conditions such that $\phi'=0$
and $a'=\pm 1$ at $a=0$. After a proper analytic continuation to the
Lorentzian region, the configuration describes an open universe
residing inside an expanding bubble.

\begin{figure}
  \begin{center}
  \leavevmode\psfig{figure=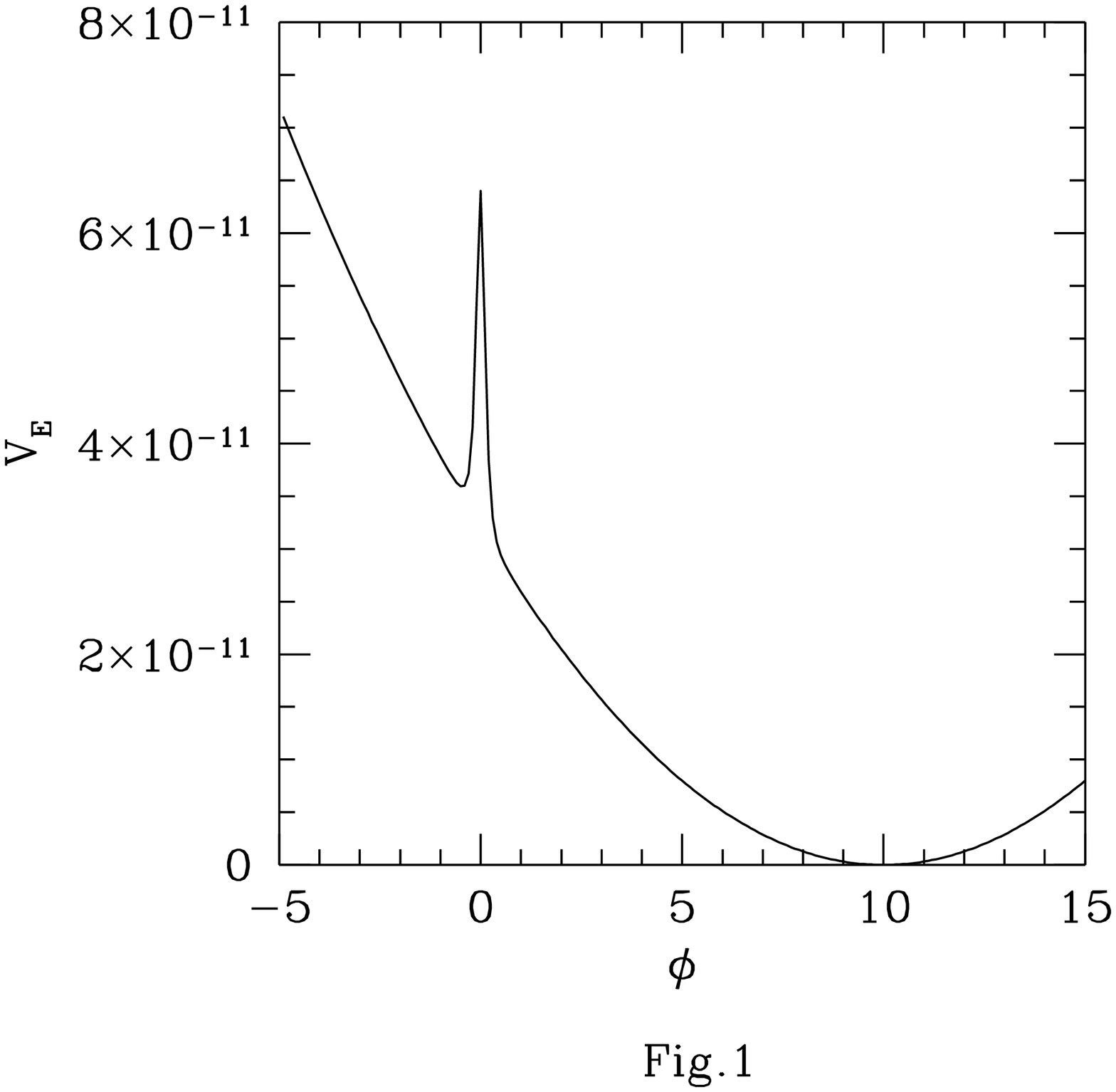,width=8.5cm}
  \end{center}
  \caption{The effective potential $V_E(\phi)$. 
    All values are normalized by $m_{pl}^*$.}
  \label{fig:fig1}
\end{figure}

\begin{figure}
  \begin{center}
  \leavevmode\psfig{figure=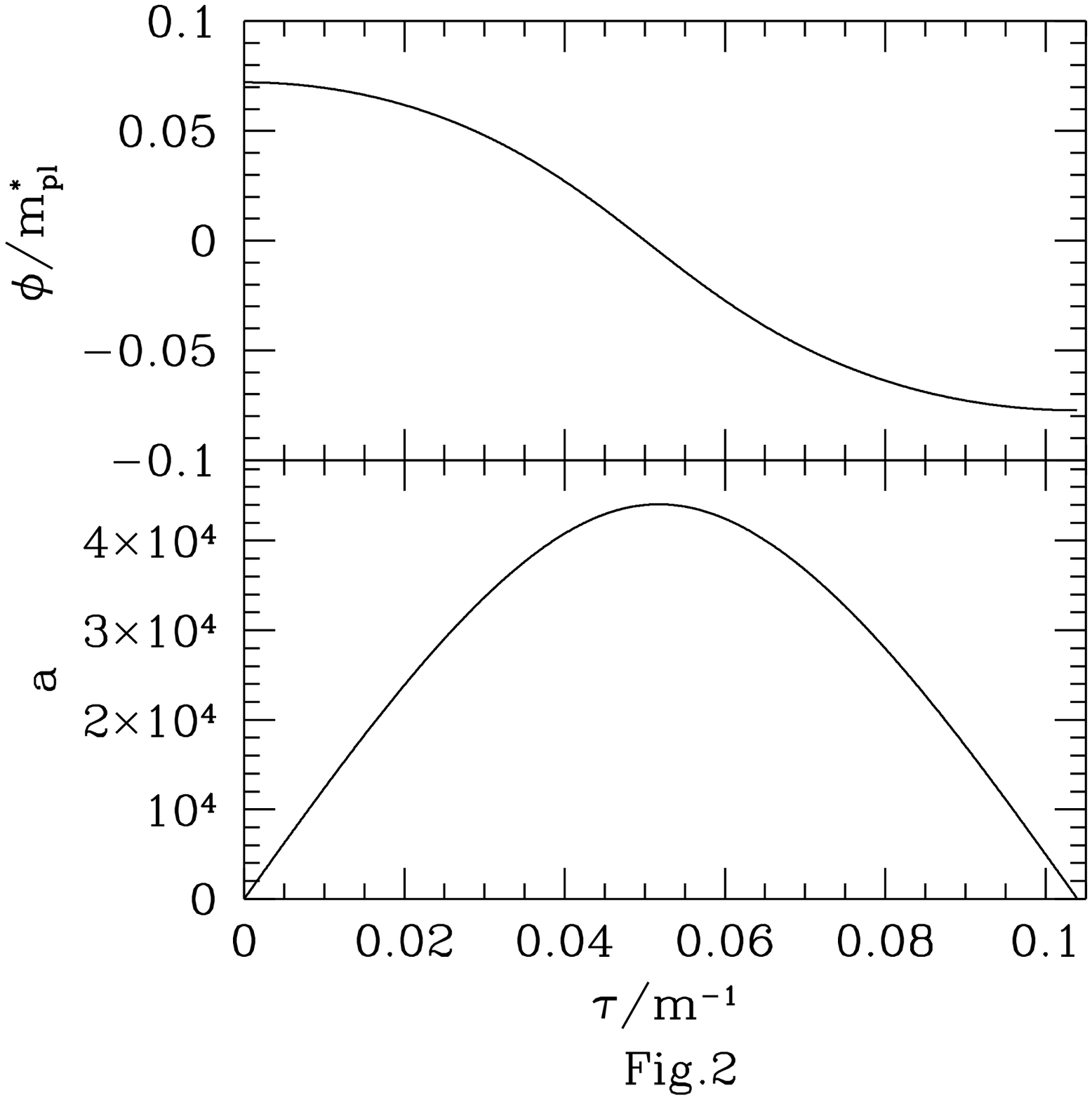,width=8.5cm}
  \end{center}
  \caption{Coleman-de Luccia instanton in our model. The upper panel
    shows the scalar field; the lower panel shows the scale factor.}
  \label{fig:fig2}
\end{figure}

As a concrete example, let us consider the following functional
forms\footnote{We note that we could in fact find a viable model (that
  is, both Coleman-de Luccia instanton and slow-roll inflation after
  the tunneling with reasonable e-fold is realized) with $V_1=0$ and
  up to quartic terms for $V_2$. But a considerable fine tuning for
  the parameters is required, and the results are not so illuminating.
  We hesitate to show them here.}
\beqa
&&f(\phi)=1+\xi\k^2\phi^2,\\
&&V_1(\phi)=V_2(\phi)={1\over 2}m^2(\phi-v)^2.
\eeqa
Clearly, this is not a unique choice. In particular, the mass scale of
$V_1(\phi)$ needs not to coincide with that of $V_2(\phi)$.  Moreover,
a different functional form for $V_2$, for example
$V_2(\phi)=\lambda(\phi^2-v^2)^2$, may be possible.  However,
$V_2(\phi)$ should have the same minimum as $V_1(\phi)$.  $f(\phi)$
should be convex and positive, and its curvature around the minimum
should be larger than that of $V_1$ and $V_2$.  For definiteness, we
choose $\xi=1$ and $v=10\times m^*_{pl}$ with $m_{pl}^{*2}\equiv
G_*^{-1}$, and then $m=8\times 10^{-7}m^*_{pl}$ for the normalization
of the density perturbations. Note that $\xi$ needs not always to be
large. The shape of the effective potential
$V_E(\phi)$ is shown in Fig.\ \ref{fig:fig1}.\footnote{We should draw
the figure using the canonically normalized scalar field $\Phi$ defined
by $\Phi=\int d\phi F(\phi)$. However, the shape does not change
qualitatively.} 
We note that this choice
of the parameters is not a unique one.  For example, another possible
choice is $\xi=100, v=1.5\times m^*_{pl}$ and $m=4\times
10^{-6}m^*_{pl}$. In Fig.\ \ref{fig:fig2}, we show the Coleman-de
Luccia instanton of our model.

\begin{figure}
  \begin{center}
  \leavevmode\psfig{figure=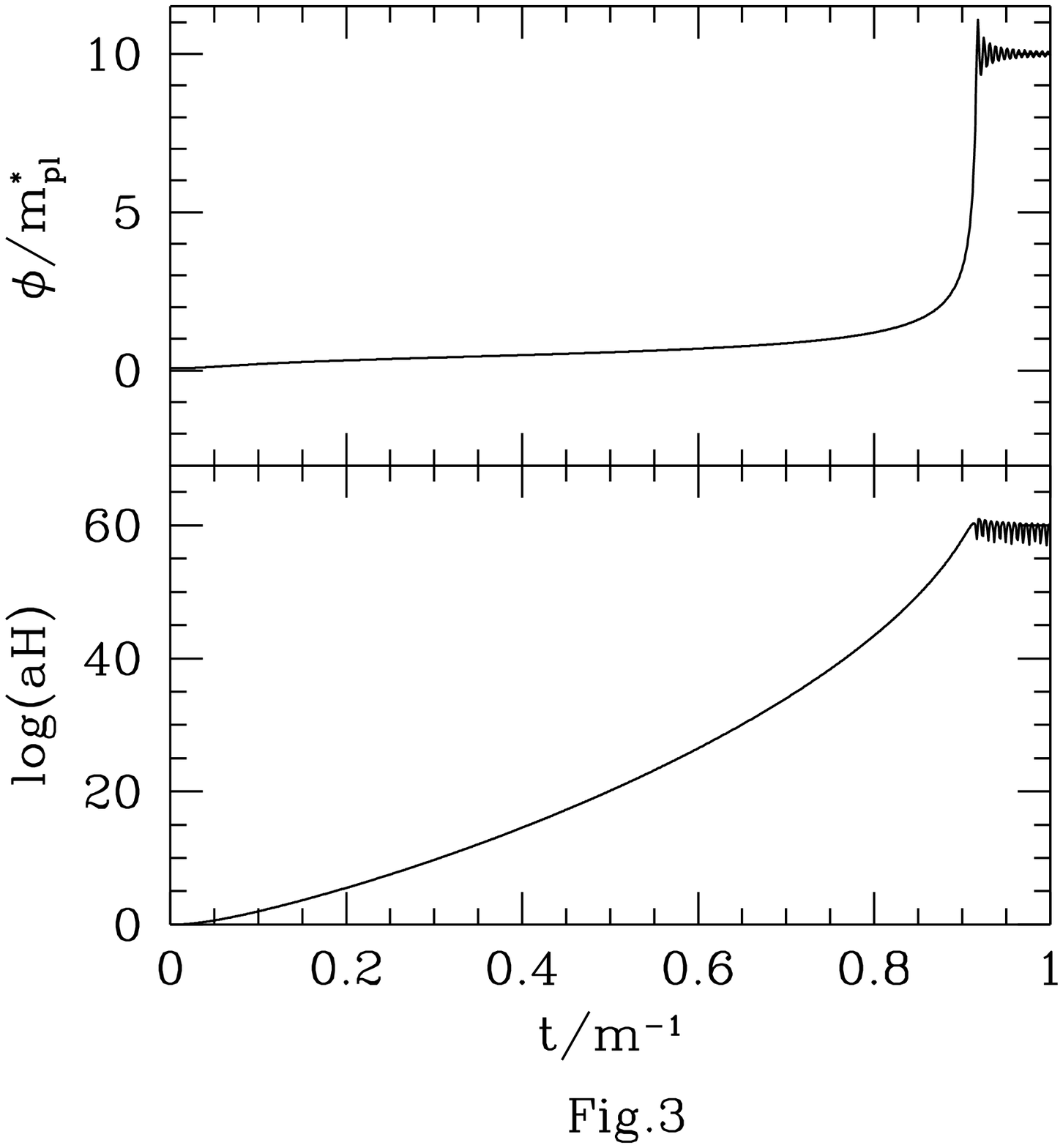,width=8.5cm}
  \end{center}
  \caption{Inflation after the tunneling. The upper panel shows the
    evolution of the scalar field. The lower panel shows the number of
    e-folds of the decrease of the comoving Hubble scale.}
  \label{fig:fig3}
\end{figure}

After the tunneling, the universe undergoes the inflationary stage.
The scale factor and the scalar field obey the following equations
after analytically continued to the Lorentzian regime
\beqa
{\ddot a\over a}&=&-{\k^2\over 3f}\left(\dot\phi^2-V\right)
-{f_{,\phi}\over 2f}{\dot a\over a}\dot\phi-{f_{,\phi\phi}\over  2f}
\dot\phi^2 -{f_{,\phi} \over 2f}\ddot\phi,\\
\ddot\phi&=&-3{\dot a\over a}\dot\phi-V_{,\phi}+{3 f_{,\phi}\over \k^2}
\left({\ddot a\over a}+\left({\dot a\over a}\right)^2-{1\over a^2}
\right),
\eeqa
where dots denote derivatives with respect to the cosmic time.  The
boundary conditions are $\dot \phi=0$ and $\dot a=0$ at $a=0$. Then,
the evolution of the scalar field and the scale factor are depicted in
Fig.\ \ref{fig:fig3}.

\begin{figure}
  \begin{center}
  \leavevmode\psfig{figure=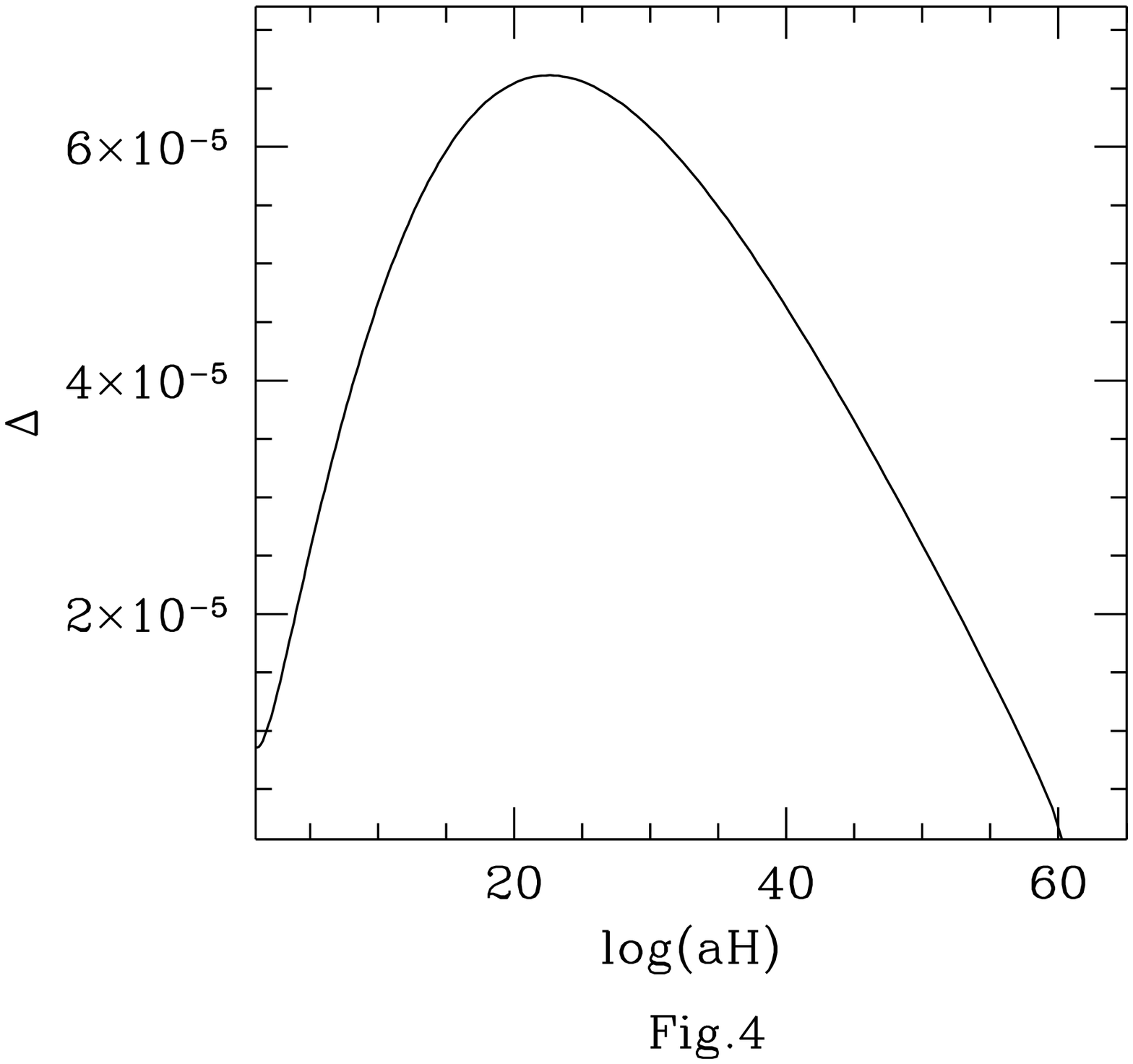,width=8.5cm}
  \end{center}
  \caption{Density perturbations $\Delta$ produced inside the bubble $N$
    e-folds after the creation of an open universe.}
  \label{fig:fig4}
\end{figure}

We calculate the function $\Delta=H^2/(5\pi\sqrt{1+6\xi}|\dot\phi|)$
which would correspond to density perturbations measured on a comoving
hypersurface in a flat matter-dominated universe\cite{ms}. Similar to
Linde's model, the amplitude of $\Delta$ has a maximum at small
$\log(aH)$ as shown in Fig.\ \ref{fig:fig4}. In our model, the
maximum is at $\log(aH)\simeq 20$.  It should be noted that the true
Planck scale $m_{pl}$ after inflation ($\phi=v$) is given by
$m_{pl}=m_{pl}^*\sqrt{1+\xi\k^2v^2} (> m_{pl}^{*})$. Therefore $\phi$
during inflation can be smaller than $m_{pl}$ although the effective
Planck scale at the beginning of inflation $m_{pl}^{eff}\equiv
m_{pl}^{*}\sqrt{1+\xi\k^2\phi^2}$ is not so much different from
$m_{pl}^{*}$.  For example, for the parameters given above, $m_{pl}
\simeq 50\times m_{pl}^{*}$ and thus $v \simeq 0.2\times m_{pl}$.

To conclude, allowing the nonminimal coupling of the inflaton to
gravity greatly expands the range of viable models for open inflation.

\acknowledgments 
We would like to thank Jun'ichi Yokoyama for useful discussion. 
This work was supported in part by JSPS Fellowship
for Young Scientists under grant No.3596~(TC) and No.4558~(MY). A part
of this work was done while one of the authors (TC) was visiting the
Aspen Center for Physics, which he acknowledges for hospitality.


\end{document}